# Square Kilometre Array: a concept design for Phase 1

**M.A. Garrett**[1-3]
[1]*ASTRON, Netherlands Institute for Radio Astronomy, Postbus 2, 7990AA, Dwingeloo, The Netherlands*
[2]*Sterrewacht Leiden, Leiden University, Postbus 9513, 2300 RA Leiden, The Netherlands.*
[3]*Centre for Astrophysics and Supercomputing, Swinburne University of Technology, Australia.*
*E-mail: :* `garrett@astron.nl`

**J.M. Cordes**
*Astronomy Department, Cornell University, Ithaca, NY 14853, USA.*

**D.R. Deboer**
*CASS, CSIRO Astronomy & Space Science, PO Box 76, Epping NSW 1710, Australia.*

**J.L. Jonas**
*Department of Physics & Electronics, Rhodes University, Grahamstown 6140, South Africa.*

**S. Rawlings**
*Department of Physics, University of Oxford Astrophysics, Keble Road, Oxford OX13RH, UK.*

**R.T. Schilizzi**
*SKA Program Development Office, The University of Manchester, Oxford Road, Manchester, UK.*

The SKA at mid and low frequencies will be constructed in two distinct phases, the first being a subset of the second. This document defines the main scientific goals and baseline technical concept for the SKA Phase 1 ($SKA_1$). The major science goals for $SKA_1$ will be to study the history and role of neutral Hydrogen in the Universe from the dark ages to the present-day, and to employ pulsars as probes of fundamental physics. The baseline technical concept of $SKA_1$ will include a sparse aperture array operating at frequencies up to 450 MHz, and an array of dishes, initially operating at frequencies up to 3 GHz but capable of 10 GHz in terms of antenna surface accuracy. An associated Advanced Instrumentation Program (AIP) allows further development of new technologies currently under investigation. Construction will take place in 2016-2019 at a total capital cost of 350M€, including an element for contingency. The cost estimates of the $SKA_1$ telescope are now the subject of a more detailed and thorough costing exercise led by the SKA Project Development Office (SPDO). The 350 M€ total for $SKA_1$ is a cost-constrained cap; an additional contingency is to reduce the overall scope of the project. The design of the $SKA_1$ is expected to evolve as the major cost estimates are refined, in particular the infrastructure costs at the two sites. The $SKA_1$ facility will represent a major step forward in terms of sensitivity, survey speed, image fidelity, temporal resolution and field-of-view. It will open up new areas of discovery space and demonstrate the science and technology underpinning the SKA Phase 2 (SKA$_2$).

## 1. Introduction

The Square Kilometre Array (SKA) is a radio telescope that is being designed to observe the sky at frequencies from 70 MHz to 25 GHz in three bands (low: 70-450 MHz, mid: 0.3-10 GHz, high: 5-25 GHz). In order to span this wide frequency range, the concept involves two main receptor types, aperture arrays (see Fig.1) and dishes with innovative feeds (see Fig.4) [1].

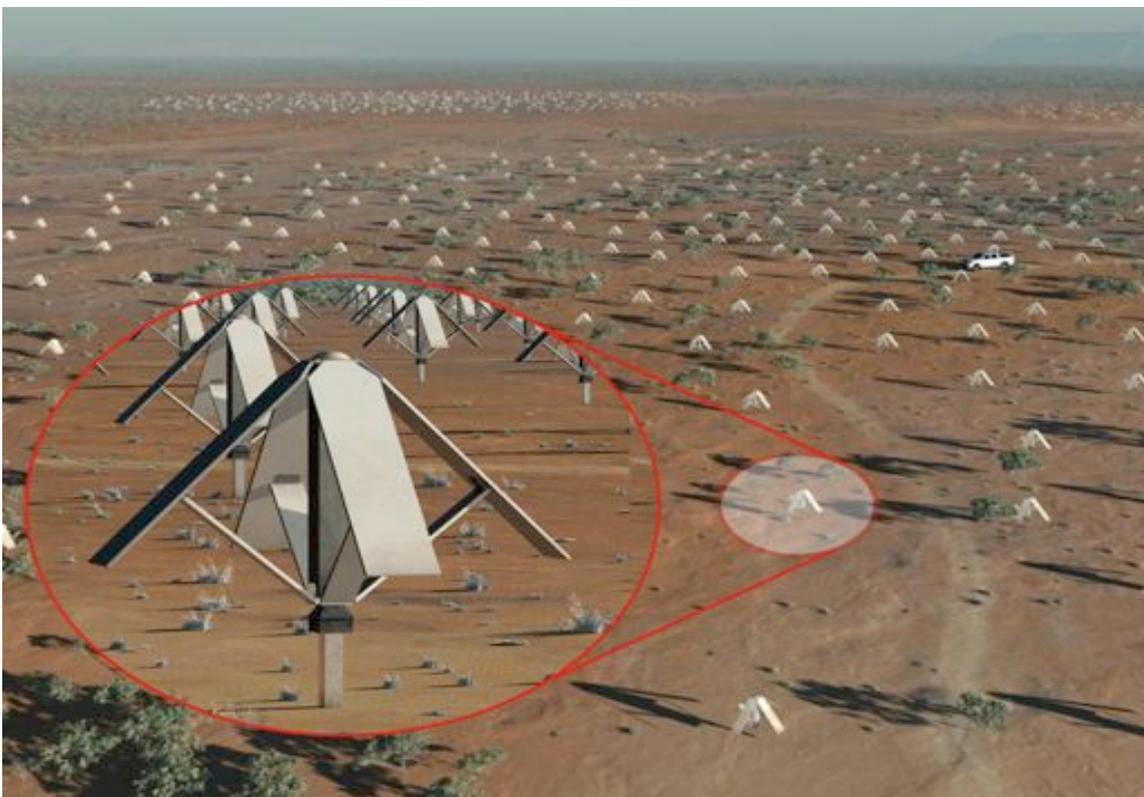

*Figure 1: The SKA$_1$ concept involves two main receptor types, sparse aperture arrays (above, courtesy of Swinburne Astronomy Productions with enhancements by MAG) & dishes (see Fig.4).*

The SKA at mid and low frequencies will be constructed first; this will take place in two distinct phases. This document defines the main scientific goals and general technical concept for the SKA Phase 1 (hereafter SKA$_1$) which is a scientific and technological subset of SKA Phase 2 (hereafter SKA$_2$). A more detailed concept design will be presented in a subsequent document. This will be essential input to the overall SKA system design and costing exercise to be completed at the end of 2012, prior to the detailed engineering design in the pre-construction phase (see timeline below).

## 2. Major Science Goals of SKA₁

The SKA Science Working Group [1, 2, 3] has identified the science topics that can be addressed by SKA$_1$. Using these papers as a guide, the SKA Science & Engineering Committee (SSEC) has identified the following major science goals that drive the technical specifications for the SKA$_1$:

(i) Understanding the history and role of neutral Hydrogen in the Universe from the dark ages to the present-day (see Fig.2), and

(ii) Detecting and timing binary pulsars and spin-stable millisecond pulsars in order to test theories of gravity (including General Relativity and quantum gravity), to discover gravitational waves from cosmological sources, and to determine the equation of state of nuclear matter (see Fig.3).

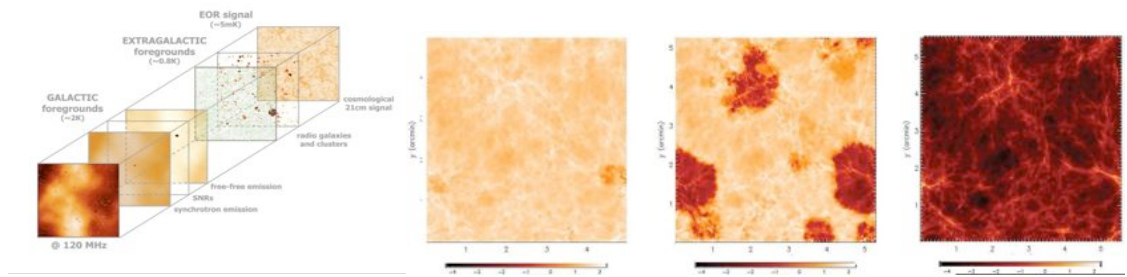

*Figure 1: Studies of the EoR by the SKA$_1$ will require a careful subtraction of the various galactic and extra-galactic foregrounds (left, courtesy Jelic et al. 2007 [4]) but the science return (right, courtesy Mellema et al. 2007 [5]) holds enormous promise.*

Addressing the themes of "Origins" and "Fundamental Physics", these two major goals are supplemented with the theme of "Discovery". In particular, the advances that SKA$_1$ represents in terms of its sensitivity, survey speed, time domain sampling and image fidelity will open up new regions of discovery space.

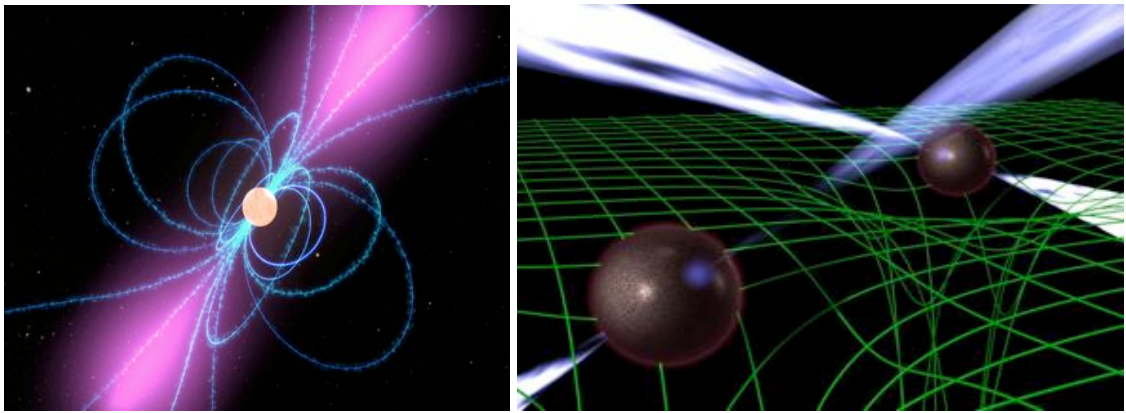

*Figure 2: Studies of pulsars (left: courtesy NASA/GSFC Conceptual Image Lab) by SKA$_1$ will test theories of gravity (right, courtesy David Champion), detect low-frequency gravitational waves, and shed light on the equation of state of nuclear matter.*

In addition to these core science drivers, SKA$_1$ will also permit us to make significant progress towards the major science goals of SKA$_2$ as a science pathfinder. A wide variety of different studies will be enabled *e.g.* detecting and imaging radio continuum emission from galaxies and active galactic nuclei to trace the evolution of galaxies, black holes, star formation and magnetism from the dawn of galaxies to the present era. Large HI galaxy surveys for cosmology and dark energy may also be conducted, in addition to transient searches (including SETI). The extra demands placed by the full SKA$_2$ science case on the SKA$_1$ system requirements, have been implicitly taken into account in the design presented here.

The SKA1 concept presented here will build on the current suite of SKA pathfinders, in particular: Apertif, ASKAP, ATA, HERA, LOFAR & MeerKAT (see elsewhere in this volume). We also note that SKA$_1$ represents the essential radio component of a suite of "Origins" and "Fundamental Physics & Discovery" instruments now being planned or under construction, including ALMA, JWST, ELTs, IXO, SPICA, CTA LIGO, LISA, LSST, Euclid/JDEM, CMBPOL, GAIA, KM3NeT and the LHC (CERN).

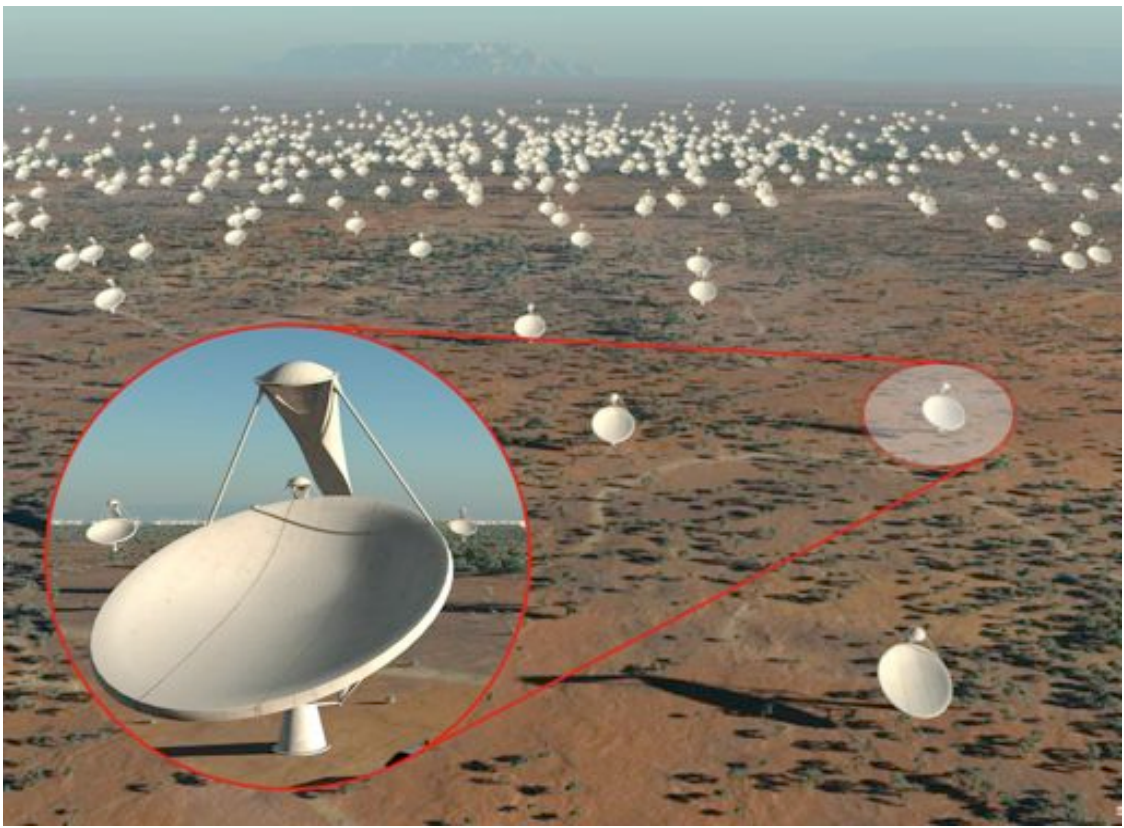

*Figure 4: The SKA$_1$ concept involves two main receptor types, aperture arrays (see Fig.1) & dishes (above, courtesy of Swinburne Astronomy Productions with enhancements by MAG).*

## 3. SKA$_1$ Technical Concept

The SKA$_1$ baseline technical concept is driven by the need to address the two major science goals identified earlier in this document *i.e.* understanding the history and role of neutral Hydrogen in the Universe from the dark ages to the present-day, and using pulsars as probes of fundamental physics.

The technologies required to achieve these two major science goals, lead us to a baseline design concept for $SKA_1$ that includes the following elements:

1) a low-frequency sparse aperture array with an $A/T_{sys}$ of up to 2000 $m^2/K$ operating at frequencies between 70 and 450 MHz. The array will be centrally condensed but some of the collecting area will be in stations located out to a maximum baseline length of 100 km from the core, and

2) a dish array with $A_{eff}/T_{sys}$ of up to 1000 $m^2/K$ using approximately two hundred and fifty 15-metre antennas, employing an instrumentation package that will use single-pixel feeds to provide high sensitivity and excellent polarisation characteristics over a frequency range of 0.45-3 GHz. The array will be centrally condensed but some of the elements will be co-located with the sparse aperture array stations out to a maximum baseline length of 100 km from the core.

The dish design will be $SKA_2$ compliant in terms of its overall performance specification, including a target r.m.s. surface accuracy of 0.5 mm (see Fig.5).

A summary of the initial set of $SKA_1$ technical specifications is presented in Table 1, at the end of this document.

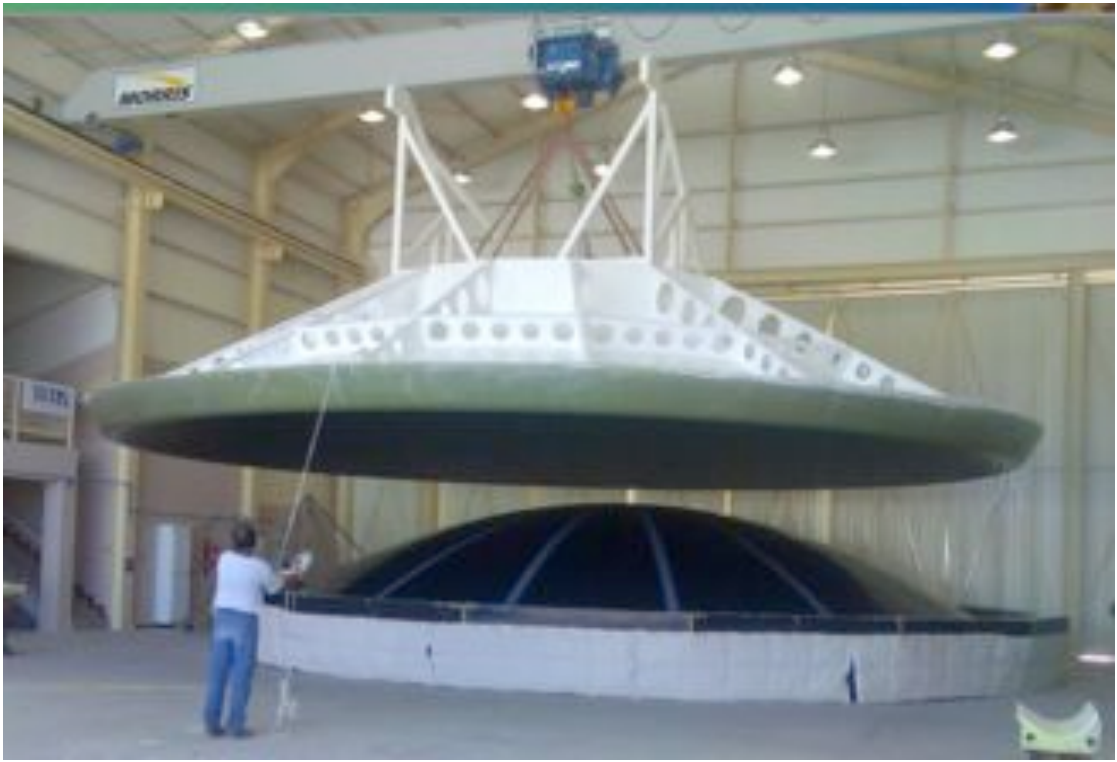

*Figure 5: Research via the US SKA Technology Development Program (TDP) and direct experience via the production of composite dishes for the South African MeerKAT project, and the metal antennas of the Australian ASKAP project, indicate that a target surface accuracy of 0.5 mm (r.m.s.) or better is achievable. Image above: courtesy of the MeerKAT project.*

## 4. SKA$_1$ Timeline

The technical timeline is as follows:

2010-12: telescope system design, prototyping and costing
2013-15: detailed engineering design & pre-construction phase
2016-19: construction, commissioning & early science observations
2016: Advanced Instrumentation Programme (AIP) decision (see below)
2020: Full SKA$_1$ science operations begin

## 5. SKA$_1$ Budget

*Pre-construction*

During the pre-construction phase from 2011 to 2015, resources at the level of 70 M€ will be required in order to: (i) cover the costs of the central SKA Project office that lead, coordinate and manage the project, and (ii) support the continued prototyping of dish and aperture array technologies, and associated sub-systems via contractual agreements with industry and in-kind contributions from the SKA partner institutes. Note that these pre-construction costs are part of the SKA Organisation budget for 2011-2015, and are therefore not included in the total cost estimates for SKA$_1$ discussed below.

*Construction*

The capital investment cost of the SKA$_1$ is capped at 350M€, including a significant element of contingency. Cost estimates of the SKA$_1$ telescope are now the subject of a detailed costing exercise led by the SKA Project Development Office (SPDO). The costs of infrastructure are expected to be a major element of the budget. Since the total cost of SKA$_1$ is capped at 350 M€, an additional contingency is to reduce the overall scope of the project. The definition of the SKA$_1$ is therefore expected to evolve as the major cost estimates are refined, in particular the infrastructure costs at the two sites.

## 6. Advanced Instrumentation Programme (AIP) for the full SKA

The SKA$_1$ baseline concept design presented here represents the first concrete step towards the realisation of the much larger and more ambitious SKA$_2$. The Advanced Instrumentation Programme (AIP) will seek to capitalise on investments made by the SKA Organisation and other parties in innovative technology development over the pre-construction period 2011-2015. In 2011-2015, advanced instrumentation systems under development for the SKA are expected to include: Phased Array Feeds (PAFs), wide-band/high frequency feeds (see Fig.6) and Dense Aperture Arrays (DAAs – see Fig.7) etc.

Given the progress expected in all of these areas of instrumentation over the next 5 years, a decision identifying the most promising system for SKA$_2$ can be made in 2016, at the start of the initial SKA$_1$ construction phase.

Advanced instrumentation such as PAFs or wideband/high frequency feeds may also be deployed as modular sub-systems on the SKA$_1$ dishes. DAAs will require the construction of a substantial standalone demonstrator. The AIP will realise an advanced system that will either greatly enhance the baseline SKA$_1$ telescope and/or will demonstrate an important technology prototype of direct relevance to SKA$_2$. The opportunity for individual research groups to fund new instrumentation relevant to the SKA should not be excluded.

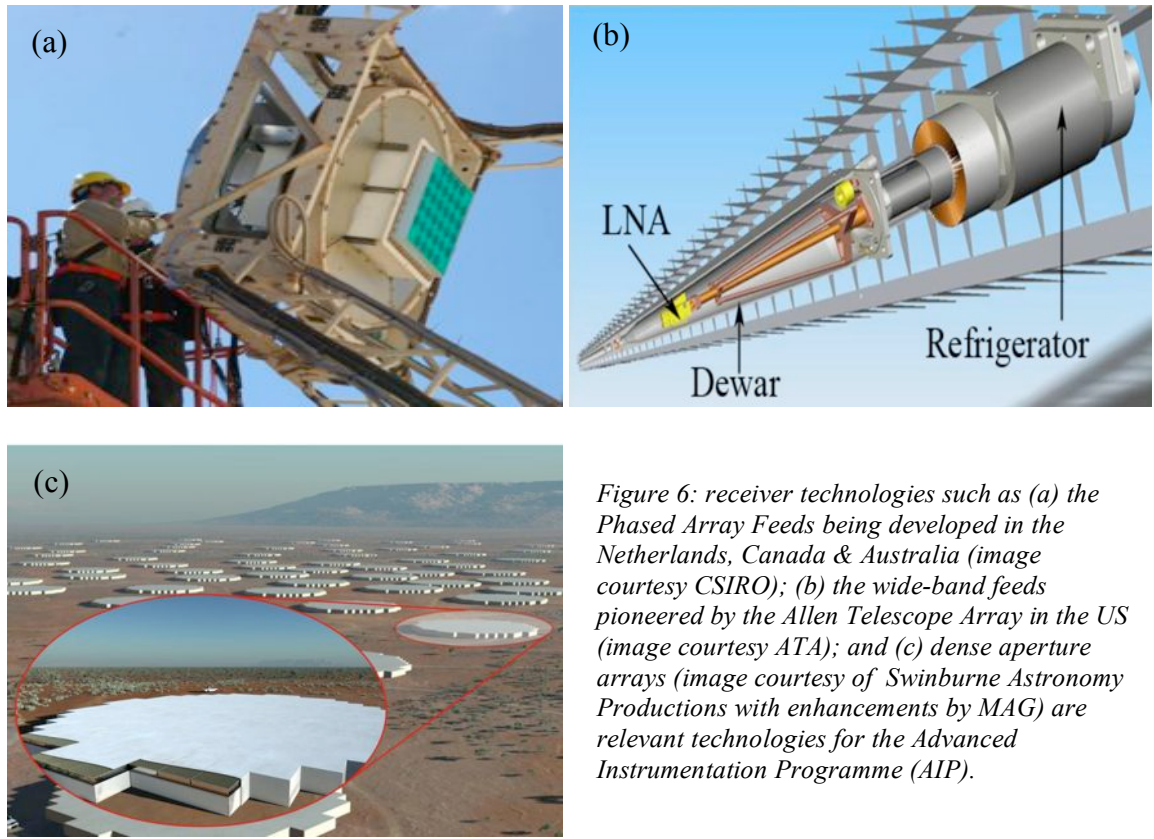

*Figure 6: receiver technologies such as (a) the Phased Array Feeds being developed in the Netherlands, Canada & Australia (image courtesy CSIRO); (b) the wide-band feeds pioneered by the Allen Telescope Array in the US (image courtesy ATA); and (c) dense aperture arrays (image courtesy of Swinburne Astronomy Productions with enhancements by MAG) are relevant technologies for the Advanced Instrumentation Programme (AIP).*

## 7. Towards SKA$_2$

The key aspect to be defined for SKA$_2$ is the relative proportion of: (i) dishes (with or without PAFs), (ii) sparse aperture arrays, and (iii) dense aperture arrays in the final system. The receptor specifications for SKA$_1$ will be developed as a result of analysing requirements for SKA$_2$, taking into account a plausible range of top-level system parameters, aligned with the relevant components of the Design Reference Mission (DRM) [3].

During the SKA$_1$ design process, components of receptors used in SKA$_1$ that are difficult or impossible to change will be designed and reviewed against the requirements of the full SKA (i.e. they will be SKA$_2$ compliant). In some, hopefully limited cases, it will be necessary for SKA$_1$ sub-systems to be replaced in order to be compatible with SKA$_2$. Certain aspects, such as the delivery of power, will require a cost-based analysis to be conducted, as to whether extensibility to SKA$_2$ is feasible.

*Acknowledgements*

The authors are members of a sub-committee appointed by the SKA Science & Engineering Committee (SSEC). The sub-committee was tasked with drafting a design concept for the SKA Phase 1 (SKA$_1$). The document presented here relies on essential input from the SSEC and its various SKA Working Groups.